\documentclass{svproc}

\usepackage[T1,hyphens]{url}

\usepackage{graphicx}
\usepackage{subfigure}
\usepackage{multirow}
\usepackage{amsmath} 
\usepackage{stmaryrd}
\usepackage{float}
\usepackage{rotating}
\usepackage{multirow}
\usepackage[titletoc, title]{appendix}
\usepackage{adjustbox}
\usepackage{bbm}
\usepackage{color}
\usepackage[table]{xcolor}
\usepackage{lscape}
\usepackage{tablefootnote}
\usepackage{subfig} 
\usepackage{caption}
\usepackage{morefloats}

\usepackage{tablefootnote}

\usepackage[utf8]{inputenc}

\usepackage{array} 
\usepackage{tabularx} 

\usepackage[misc,geometry]{ifsym}

\usepackage{comment}

\usepackage[square,numbers]{natbib} 
\makeatletter
\renewcommand\bibsection%
{
 \section*{\refname
 \@mkboth{\MakeUppercase{\refname}}{\MakeUppercase{\refname}}}
}
\makeatother

\usepackage{stfloats}

\usepackage{todonotes}

\usepackage{multicol,booktabs}

\usepackage{navigator}

\begin{document}

\title{Ontology engineering to model the European cultural heritage: The case of \emph{Cultural gems}$^{*}$}

\titlerunning{Ontology engineering for Cultural gems}

%
%
\author{Valentina Alberti\inst{1} \and Cinzia Cocco\inst{1} \and Sergio Consoli\inst{1}\Letter \and Valentina Montalto\inst{2} \and Francesco Panella\inst{1} }

%
\authorrunning{Alberti et al.} 
%
\institute{European Commission, Joint Research Centre (DG JRC), Ispra (VA), Italy. 
\email{ [name.surname]@ec.europa.eu}
\\
\vspace{2 mm}$^*$Authors listed in alphabetic order.
}

\maketitle 

\begin{abstract}

\emph{Cultural gems} is a web application conceived by the European Commission's Joint Research Centre (DG JRC), which aims at engaging people and organisations across Europe to create a unique repository of cultural and creative places. 
The main goal is to provide a vision of European culture in order to strengthen a sense of identity within a single European cultural realm.
\emph{Cultural gems} maps more than 130,000 physical places in over 300 European cities and towns, and since 2020 it also lists online cultural initiatives. 
The new release 
aims, among other, to increase the interoperability of the application. 
At this purpose, we provide an overview on the current development 
of an ontology for \emph{Cultural gems} used to map cultural heritage in European cities
by using Linked Open Data (LOD) standards, and making the data \emph{FAIR}, that is \emph{F}indable, \emph{A}ccessible, \emph{I}nteroperable, and \emph{R}eusable. 
We provide an overview of the methodology, 
presenting the structure of the ontology, and the services and tools we are currently building on top. 
\end{abstract}

\section{Introduction and background} \label{introduction} 

Culture can be considered one of the ways to preserve and re-launch cities' attractiveness in the face of sanitary crisis \citep{Agostino2020362,Agostino20201}. People dwelling in cities can join forces, supporting the process, sharing information on culture and creativity, and highlighting what is unique in each city.
\emph{Cultural gems} (CG)\footnote{\emph{Cultural gems} application: \url{https://culturalgems.jrc.ec.europa.eu/}} is a crowdsourced web application created by the Joint Research Centre (DG JRC)\footnote{\url{https://joint-research-centre.ec.europa.eu/index_en}} of the European Commission.  
The platform, which is free and open source, has the objective of mapping relevant artistic and cultural locations in cities of the Euro area.
The main goal is to document the range of culture and creativity found in European cities and towns by producing crowdsourced maps and a shared database of these locations.
The source of information on the set of cultural locations comes from both OpenStreetMap\footnote{
\url{https://www.openstreetmap.org/}} and the data shared by municipalities, research organizations, and other private and public institutions in Europe.
In this way users are presented with enriched maps of EU cities, which are easily sharable and visualizable in user-friendly ways.
The application offers a digital tool and resources to help local authorities and individuals which work, or are just interested, in the artistic and cultural domains, in order to promote these sectors within their cities \citep{Mudge2007}.



The platform was first inaugurated in December 2018 in the context of the European Year of Cultural Heritage\footnote{
\url{https://culture.ec.europa.eu/cultural-heritage/eu-policy-for-cultural-heritage/european-year-of-cultural-heritage-2018}}.
\emph{Cultural gems} has been featured in several policy documents, such as: \emph{The European framework for action on cultural heritage}, coming from the 
2018 European Year of Cultural Heritage; \emph{Tourism and transport in 2020 and beyond}\footnote{
\url{https://eur-lex.europa.eu/legal-content/EN/TXT/?uri=CELEX:52020DC0550}}, as a tool to support proximity tourism (COM/2020/550); \emph{EU guidelines for the safe resumption of activities in the cultural and creative sectors - COVID-19}, as one of the actions to support the sustainable recovery of the cultural sectors (2021/C 262/01).\footnote{
\url{https://eur-lex.europa.eu/legal-content/EN/TXT/?uri=CELEX:52021XC0705(01)}} 
Since its launch, CG is continuously evolving to meet users' needs. Currently, the application contains information on more than 130,000 cultural venues 
in over 300 European cities. 
\emph{Cultural gems} has been also used as a communication and outreach tool, highlighting cultural initiatives both at European and local level. Since 2020 a dedicated section lists more than 400 cultural initiatives accessible online.\footnote{\url{https://culturalgems.jrc.ec.europa.eu/eu-culture-from-home}} In 2021, \emph{Cultural gems} was enriched by contributions and collaborations with local authorities, universities, schools and users from all over Europe. 
In some cities contributions were particularly detailed and well-finished, such as in \jumplink{https://culturalgems.jrc.ec.europa.eu/map/143905}{Izola} (Slovenia), \jumplink{https://culturalgems.jrc.ec.europa.eu/map/40}{Karlsruhe} (Germany), \jumplink{https://culturalgems.jrc.ec.europa.eu/map/14}{Plovdiv} (Bulgaria), \jumplink{https://culturalgems.jrc.ec.europa.eu/map/148}{Lisbon}, and \jumplink{https://culturalgems.jrc.ec.europa.eu/map/150}{Coimbra} (both in Portugal). Further, in occasion of the Portuguese Presidency of the Council of the European Union, ad-hoc activities were led with Portuguese cities that have been particularly active as contributors. A dedicated page presents the main contributions to the Portuguese cities.\footnote{\url{https://culturalgems.jrc.ec.europa.eu/portuguese-semester}} The section ``EU Culture from Home'' kept growing, supporting users during the confinement period due to the Covid-19 pandemic, while more and more city stories were used to map intangible heritage in European cities.\footnote{As an example of a city story describing a tradition of the 
city, see \jumplink{https://culturalgems.jrc.ec.europa.eu/map/148/27213/139652}{Fado in Lisbon}.} 
In 2022, the platform is changing data infrastructure to improve data interoperability and information accessibility. 
At the purpose, we want to exploit the enormous potential of Artificial Intelligence for cultural heritage \citep{Fiorucci2020102,Diaz-Rodriguez2020317,EgarterVigl2021673,Mudge2007}, and in particular those derived by the adoption of 
Semantic technologies and Linked Open Data (LOD) \citep{Heath20111,Hyvnen2020187}, to facilitate the interconnection with disparate datasets and ontologies in the cultural heritage field \citep{Stasinopoulou2007165,Popovici2011105}, such as Europeana\footnote{
\url{https://www.europeana.eu/en}} \citep{europeana} and the Knowledge Graph of the Italian Cultural Heritage (ArCo)\footnote{
\url{http://dati.beniculturali.it/arco/}} 
\citep{Carriero201936,arcoJournal}, using metadata standards, and increasing semantic interoperability of the application.



\section{\emph{Cultural gems} ontology} \label{Model} 

An ontology for \emph{Cultural gems} has been designed from the main 
CG classes used in the application, whose current classification is loosely based on the ``concentric circles model of cultural industries'' by \citet{Throsby2008}. 
The ontology aims at modelling 
this cultural heritage data. 
The categories are mainly organised to match also the OpenStreetMap categorization\footnote{
\url{https://wiki.openstreetmap.org/wiki/Map_features}} relevant 
to our mapping purposes for both interoperability and clarity purposes. 

The goal consists of building an ontology that is compatible, and aligned whenever possible, with existing ontologies in the related domain, that are used as a de facto standard for representing cultural heritage data. These include, for example, the already mentioned Europeana and ArCo ontologies, and services like the Hellenic Aggregator of Digital Cultural Content\footnote{
\url{https://www.searchculture.gr/aggregator/portal/}}, among others, for linking and aggregating the generated application data with the various cultural heritage LOD available online. 

We heavily rely on ontology design patterns (ODPs) principles \citep{Gangemi2005262} to build our ontology, that is, whenever possible we reuse existing ODPs from online ontology repositories. Reused patterns are annotated with the OPLa ontology \citep{Shimizu201823,Asprino2021} in order to facilitate ontology class mapping and identification. For example, we directly reuse classes and properties from the available OntoPia 
Public Administration vocabulary\footnote{
\url{https://github.com/italia/daf-ontologie-vocabolari-controllati/tree/master/}}
%
and from an ontology describing cultural events and sites (Cultural-ON ontology)\footnote{
\url{https://dati.beniculturali.it/cultural-ON/ENG.html}} \citep{Lodi2017}. 
We indirectly reuse patterns from existing ontologies, e.g. CIDOC-CRM\footnote{
\url{http://www.cidoc-crm.org/}}, and include explicit alignments to them. 
Furthermore, our ontology definition reuses various classes and properties of the ArCo network of ontologies, connected by \emph{owl:imports} axioms.
The main classes of our \emph{Cultural gems} ontology are mapped to RDF/OWL as subclasses of the top-level hierarchy of ArCo, 
in particular: \emph{:CulturalProperty}, which has two subclasses \emph{:TangibleCulturalProperty} and \emph{:IntangibleCulturalProperty}. The first is further specialized
in \emph{:MovableCulturalProperty} and \emph{:ImmovableCulturalProperty}. Other specific types of cultural properties we reuse are: \emph{:ArchaeologicalProperty}, \emph{:HistoricOrArtisticProperty},  and \emph{:MusicHeritage}. The cultural events module, which extends the Cultural-ON ontology, has been also used to map cultural events and exhibitions involving a cultural property. 
\begin{figure}[b!]
\begin{center}
 \includegraphics[width=1.00\linewidth]{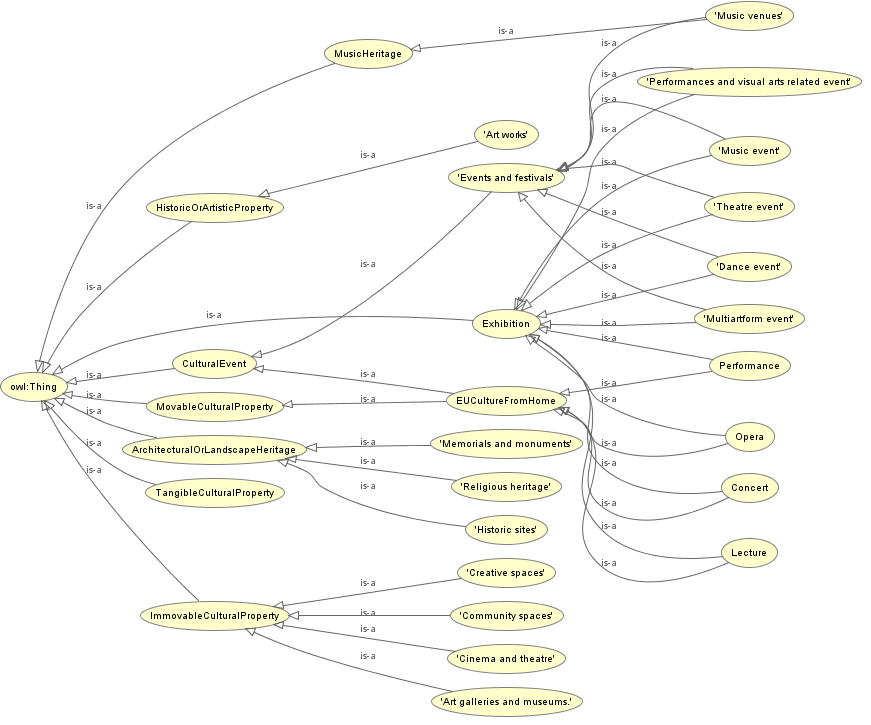}
	\caption{Hierarchy of the main classes of the \emph{Cultural gems} ontology.}
	\label{onto-mapping-arco}
\end{center}
\end{figure}

The main classes of \emph{Cultural gems} are then mapped to those class definitions, as illustrated in Figure \ref{onto-mapping-arco}. The top CG hierarchy consists of the following classes: 
\begin{itemize}
    \item[$\rightarrow$] \emph{:EUCultureFromHome} - Class representing cultural initiatives in European cities accessible online. For example, travel restrictions and social distancing might be limiting the possibility to visit venues and to taste cultural fragrances of European cities and towns in person. Museums, theatres, local cultural organisations, libraries, and many more work to keep culture alive online in difficult times. This category maps a selection of initiatives and organised events accessable online.
    \item[$\rightarrow$]\emph{:Cinemas and theatres} - Class representing cultural venues such as cinemas, theaters or opera houses. 
    \item[$\rightarrow$] \emph{:Art galleries and museums} - Class representing cultural venues such as art galleries and museums.
    \item[$\rightarrow$] \emph{:Artworks} - Class representing public-space artworks created and displayed outside of the typical art gallery setting.
    \item[$\rightarrow$] \emph{:Creative spaces} - Category that depicts physical objects and components at various scales that are intended to encourage or support creative business workflows and creativity.
    \item[$\rightarrow$] \emph{:Historic sites} - Set of official locations where historical artifacts from the realms of, among others, politics, war, culture, or society, have been preserved because of their cultural heritage value. Historic sites 
    are frequently covered by legal protection, and many of them have received formal national historic site designation. Any building, location, landascape, or structure that is locally, regionally, or nationally significant, qualifies to be classied as a historic site. This often implies that the location must be at least 50 years old.
    \item[$\rightarrow$] \emph{:Religious heritage} - Any type of property with religious or spiritual connotations, such as, among others, churches, sanctuaries, cemeteries, etc., falls under the category of religious heritage venues.
    \item[$\rightarrow$] \emph{:Memorials and monuments} - Class that represents various forms of historical monuments and attractions. 
    \item[$\rightarrow$] \emph{:Events and festivals} - Class representing various forms of events and festivals.
    \item[$\rightarrow$] \emph{:Music venues} - Class that represents any venue used for a concert or musical performance, including recording and rehearsal studios.
    \item[$\rightarrow$] \emph{:Community spaces} - Class representing various kinds of community spaces.
\end{itemize}

Note also that the location module of ArCo has been leveraged to represent spatial and geometry data. A cultural gem in the application may be assigned with multiple locations, which are represented by means of the ontology class \emph{a-loc:LocationType}. Furthermore, it can be the case that a cultural location of a gem is valid only within a specific time interval. This is represented by using the \emph{a-loc:TimeIndexedTypeLocation} ontology class, that is an extension of the TimeIndexedSituation ontology pattern.\footnote{
\url{http://www.ontologydesignpatterns.org/cp/owl/timeindexedsituation.owl}} The reference namespace for the ontology definition is: \textbf{https://culturalgems.jrc.ec.europa.eu/ ontology/cultural-gems/}, accounting so far to an overall of 67 classes. 

\emph{Cultural gems} data in the application are then mapped as individuals of the CG ontology by means of an ETL (Extract-Transform-Load) Python job performed daily (at night).
The resulting data ontology, available in both 
Turtle and RDF/XML formats, currently accounts to around 2.9M triples. The reference namespace for the data 
is: \textbf{https://culturalgems.jrc.ec.europa.eu/resource/}. 

Currently, data resources have been linked to DBpedia\footnote{
\url{https://www.dbpedia.org/}} and GeoNames\footnote{
\url{http://www.geonames.org/}} using \textbf{owl:sameAs} axioms. Further alignments to other popular cultural heritage ontologies, 
such as Europeana and ArCo data, will be provided soon. This entity linking task, object of currently on-going work, is performed by using the LIMES tool\footnote{
\url{https://aksw.org/Projects/LIMES.html}}, 
a widely-employed entity linking and discovery tool for linked data.


Commonly-used style guidelines for labeling and representing ontology definitions and resources have been employed in our ontology engineering exercise. In particular, ontology data names have been represented in lowercase, substituting eventual space characters with dashes. Ontology definition class names, instead, have been represented using uppercase. 
Source ontology definition and data files are available in RDF/XML and Turtle formats within the Joint Research Centre Data Catalogue\footnote{Joint Research Centre Data Catalogue: \url{https://data.jrc.ec.europa.eu/}} at the following permanent location:
\url{https://data.jrc.ec.europa.eu/dataset/9ee32efe-af81-48e4-8ad6-a0db06802e03}. These ontologies will be officially released soon also within the European Data portal\footnote{European Data portal: \url{https://data.europa.eu/en}}, the official data repository for European data.

\section{Interaction and visualization of the data} \label{expe}

Currently we are implementing various intuitive and user-friendly access services
to the designed ontologies, 
including content negotiation, visualization and navigation of data, improved exploitation of the available information and knowledge discovery.
%
We enable 
the interested community 
to consume the produced data and ontology under Creative Commons Attribution 3.0 Unported (CC BY 3.0) license\footnote{\url{https://creativecommons.org/licenses/by/3.0/}}. 
Ontology resources 
are stored in the RDF named-graph: \jumplink{https://jeodpp.jrc.ec.europa.eu/ftp/jrc-opendata/CC-COIN/cultural-gems/ontology-individuals/cultural-gems-resources.owl}{https://culturalgems.jrc.ec.europa.eu/resource/}, while the ontology definition
in: 
\jumplink{https://jeodpp.jrc.ec.europa.eu/ftp/jrc-opendata/CC-COIN/cultural-gems/ontology-definition/cultural-gems.owl}{https://culturalgems.jrc.ec.europa.eu/ontology/cultural-gems/}.

Technically, the access to the data and ontology is possible by means of SPARQL queries
on 
CELLAR\footnote{\url{https://op.europa.eu/en/web/webtools/linked-data-and-sparql}}, the Publications Office's common repository of metadata and content, and using its REST APIs SPARQL endpoint services.\footnote{\url{https://data.europa.eu/data/datasets/sparql-cellar-of-the-publications-office}}
SPARQL is referred by W3C as the standard language for referencing RDF data and interacting with it. The SPARQL query language interface has a text box where queries can be entered. 

The REST web service to access the dedicated CELLAR SPARQL endpoint\footnote{
\url{https://publications.europa.eu/webapi/rdf/sparql}} 
requires the user to specify as input a given SPARQL query, while giving the result of the query as output in one of the formats that follows: \emph{text/html},
\emph{text/rdf +n3}, \emph{application/xml},
\emph{application/json}, or \emph{application/rdf+xml}. 
Suppose, as an example, that you want to get all the RDF data from the CELLAR Sparql enpoint about the \jumplink{https://culturalgems.jrc.ec.europa.eu/map/148/27213}{Museu do Fado} gem, whose resource in the data ontology corresponds to \textbf{https://culturalgems.jrc.ec.europa.eu/resource/cultural-gems/27213}.\\  This would be translated in the Sparql query:\\

\noindent
\textit{DESCRIBE $<$https://culturalgems.jrc.ec.europa.eu/resource/cultural-gems/27213$>$}\\

\noindent
that, if executed in CELLAR, would produce the result available at \url{https://publications.europa.eu/webapi/rdf/sparql?default-graph-uri=&query=DESCRIBE+%3Chttps%3A%2F%2Fculturalgems.jrc.ec.europa.eu%2Fresource%2Fcultural-gems%2F27213%3E&format=text%2Fhtml&timeout=0&debug=on&run=+Run+Query+}.

\begin{figure}[b!]
\begin{center}
    \fboxsep=0.5mm
    \fboxrule=0.5pt
    \fcolorbox{black}{white}{\includegraphics[width=0.95\linewidth]{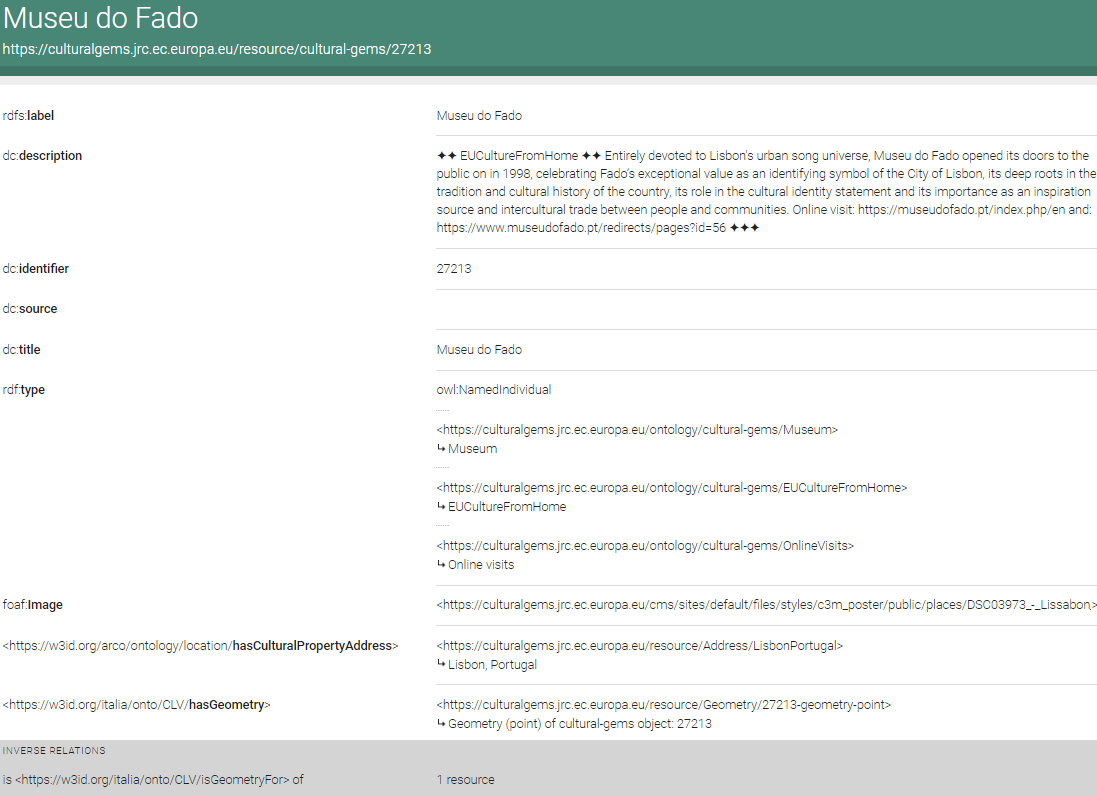}}
	\caption{LodView visualization of the Museu do Fado gem.}
	\label{lodview-museum-fado}
\end{center}
\end{figure}


We also adopt the \emph{Live OWL Documentation Environment
(LODE)} to browse the ontology in an human-readable way. LODE indeed enable the visualization in intuitive HTML pages of the ontology general axioms, namespace declarations, named individuals, classes and, finally, data, annotation and object properties.\footnote{\url{http://ec2-54-154-163-28.eu-west-1.compute.amazonaws.com:8090/lode/extract?url=https://jeodpp.jrc.ec.europa.eu/ftp/jrc-opendata/CC-COIN/cultural-gems/ontology-definition/cultural-gems.owl}}
The entire ontology is also accessible as a force-directed graph
visualization through \emph{WebVOWL}.
\footnote{\url{https://service.tib.eu/webvowl/#iri=https://jeodpp.jrc.ec.europa.eu/ftp/jrc-opendata/CC-COIN/cultural-gems/ontology-definition/cultural-gems-skeleton.owl}}
The interaction with these tools enable customization of the visualizations so that the user is presented with a user-friendly description of the ontology elements, and is able to exploit better the underlying information.

\begin{figure}[b!]
\begin{center}
    \fboxsep=0.5mm
    \fboxrule=0.5pt
    \fcolorbox{black}{white}{\includegraphics[width=0.90\linewidth]{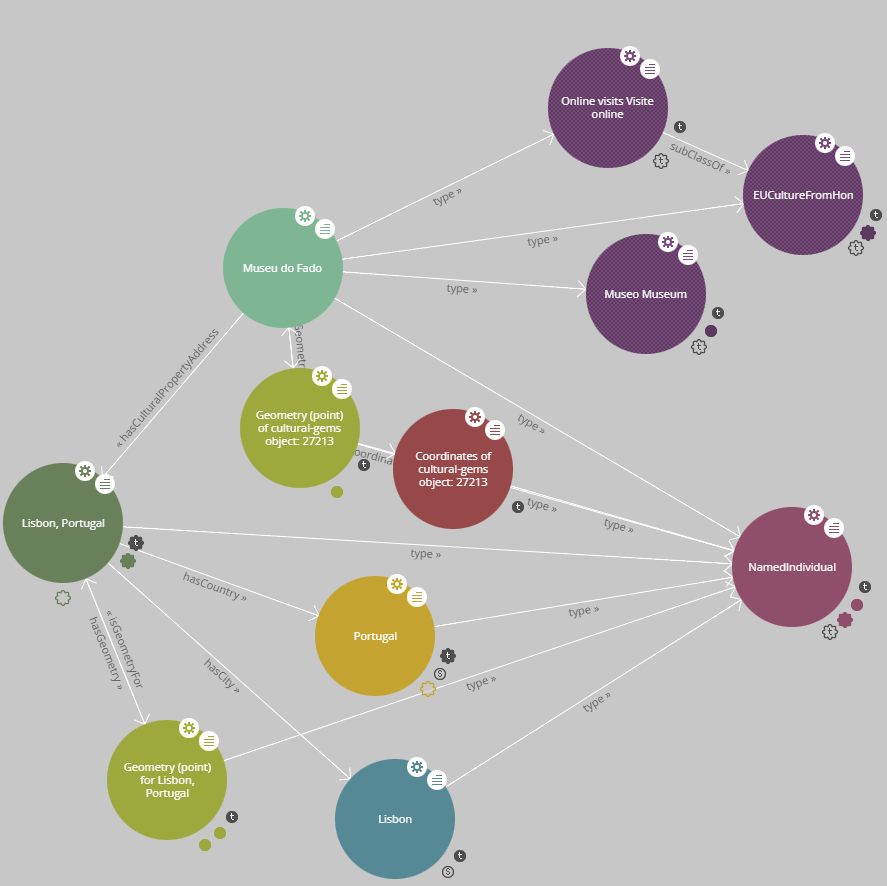}}
	\caption{View of the relationships of the Museu do Fado gem to the other data entities using LodLive.}
	\label{lodlive-museum-fado}
\end{center}
\end{figure}

We are also integrating two further visualization tools: LodView and LodLive. 
LodView\footnote{
\url{http://lodview.it/}} is a web application developed in the Java language which provides dereferentiation of URIs following W3C standards.
It basically supports users by providing representations of our RDF resources through custom intuitive HTML pages. The tool implements content negotiation of our ontology data, and allows to download the selected RDF resource in various formats, such as \textit{xml}, \textit{ntriples}, \textit{turtle}, and \textit{ld+json}. 

For example, the LodView representation of the \jumplink{https://culturalgems.jrc.ec.europa.eu/map/148/27213}{Museu do Fado} gem, in Portugal, is available at \url{http://ec2-54-154-163-28.eu-west-1.compute.amazonaws.com:441/lodview/resource/cultural-gems/27213.html} (see also Figure \ref{lodview-museum-fado}). Here you can exploit and navigate the data entity, seeing for example that this gem belongs both to the \jumplink{http://ec2-54-154-163-28.eu-west-1.compute.amazonaws.com:441/lodview/ontology/cultural-gems/Museum.html}{Museum} and \jumplink{http://ec2-54-154-163-28.eu-west-1.compute.amazonaws.com:441/lodview/ontology/cultural-gems/EUCultureFromHome.html}{EU Culture from Home} categories, and downloading the raw data in one of the available formats, e.g. json: \url{https://publications.europa.eu/webapi/rdf/sparql?output=application%2Fodata%2Bjson&query=DESCRIBE+%3Chttps://culturalgems.jrc.ec.europa.eu/resource/cultural-gems/27213%3E}, or csv: \url{https://publications.europa.eu/webapi/rdf/sparql?output=text%2Fcsv&query=DESCRIBE+%3Chttps://culturalgems.jrc.ec.europa.eu/resource/cultural-gems/27213%3E}, among others.

The second tool, LodLive,\footnote{
\url{http://lodlive.it/}} allows to depict the RDF data via an effective graph representation, enabling the navigation of the ontology resources. The user is able, for instance, to expand the relationships of a given ontology resource in an automated way and explore the structure of the RDF data. It is also possible to associate images and geo-coordinates to the data instances, and evaluate \textbf{owl:sameAs} and inverse relations.
Figure \ref{lodlive-museum-fado} shows, for instance, the relations of the \jumplink{https://culturalgems.jrc.ec.europa.eu/map/148/27213}{Museu do Fado} gem to the other data entities using the LodLive tool.

We plan to integrate further semantic services along with content negotiation of the data into the production environment of the application soon. 

\section{Conclusion} \label{conclusions} 

We have described our work 
related to the development of an ontology for the \emph{Cultural gems} web application, which supports 
on mapping relevant artistic and cultural locations in cities of the Euro
area.
We expect to achieve various added values to the current application.
Given that the adoption of ontologies allows a smooth linking of heterogenous data sources, the platform will certainly benefit on an increased flexibility on data integration, enabling semantic interoperability.
Data quality will also improve since the (re)adoption of existing ontologies forces to implement 
mechanisms and procedures to clean the underlying information and correct progressively errors within the data layer of the application.
New services to end-users will also be offered to both the private and public sectors, as a result of the larger availability of information to easily consume from the platform.
Costs are also reduced, given that the reuse of existing ontologies in the platform 
brings to lower the costs related to software development and mantainance services.
Our final goal consists in further promoting interconnections among practitioners and researches involved in the cultural heritage field into a common dialogue, contributing to the EU-wide project of culture and creativity promotion.

%
%
 %




\section*{Acknowledgements}
The authors would like to thank the CELLAR team at the Publications Office of the European Union for their support on the use of their public knowledgebase for hosting our ontology and interacting with it.



%
%
\bibliographystyle{abbrvnat}

\bibliography{bibliofile}

\end{document}